# Collisionless wave dissipation in a D-shaped tokamak with Solov'ev type equilibrium

**N.I. Grishanov[1], A.F.D. Loula[1], A.L. Madureira[1], J. Pereira Neto[2]**

[1]*Laboratório Nacional de Computação Científica, Petrópolis, RJ, BRASIL*
[2]*Universidade do Estado do Rio de Janeiro, Rio de Janeiro, RJ, BRASIL*

**Abstract:** Parallel permittivity elements are derived for RF waves in an axisymmetric D-shaped tokamak with Solov'ev type equilibrium. Drift-kinetic equation is solved separately for untrapped (passing or circulating) and three groups of trapped particles as a boundary-value problem, accounting for the bounce resonances. A coordinate system with the "straight" magnetic field lines is used. Our dielectric permittivity elements are suitable to estimate the wave dissipation by electron Landau damping (e.g., during the plasma heating and current drive generation) in the frequency range of the Alfvén and fast magnetosonic waves, for both the large and low aspect ratio tokamaks with circular, elliptic and D-shaped magnetic surfaces. Dissipated wave power is expressed by the summation of terms including the imaginary parts of both the diagonal and non-diagonal elements of the parallel permittivity.

## 1. Introduction

Spherical Tokamaks (or Low Aspect Ratio Tokamaks) represent a promising route to magnetic thermonuclear fusion [1-6]. To achieve fusion conditions in these devices additional plasma heating must be employed. Effective schemes of heating and current drive in tokamaks can be realised using the collisionless dissipation of RF waves (e.g., Alfvén, Fast Magnetosonic and Lower Hybrid Waves) by electron Landau damping, transit time magnetic pumping (TTMP), cyclotron and bounce wave-particle interactions, etc.

Kinetic wave theory in any toroidal plasma should be based on the solution of Vlasov-Maxwell's equations [7-15]. However, this problem is not simple even in the scope of the linear theory since to solve the differential wave (or Maxwell's) equations one should use the complicated integral dielectric characteristics valid in the given frequency range for realistic two- or three-dimensional plasma models. The form of the dielectric (or wave conductivity) tensor components, $\varepsilon_{ik}$, depends substantially on the geometry of an equilibrium magnetic field and, accordingly, on the chosen geometrical coordinates.

The specific toroidal effects arise since the parallel velocity of plasma particles in tokamaks is not constant, $v_\parallel = \mathbf{v} \cdot \mathbf{h} \neq const$. As a result,

- depending on the pitch angle the plasma particles should be split in the populations of the trapped *(t)* and untrapped *(u)* particles;
- trajectories of *t*- and *u*-particles are different;
- Vlasov equation should be resolved separately for each particle group;
- the *t*- and *u*-particles give different contributions to $\varepsilon_{ik}$;
- the Cherenkov-resonance conditions of the *t*- and *u*-particles are different;
- wave dissipation by the *u*- and *t*-particles depends on the ratio of $v_{ph}/v_T$ and $\omega/\omega_b$;
- another interesting feature in tokamaks is the contribution of all spectrum of *E*-field to the *m*-th harmonic of the perturbed current density: $4\pi i j_i^m / \omega = \sum_{m'}^{\pm\infty} \varepsilon_{ik}^{m,m'} E_k^{m'}$.

All these features of the Large Aspect Ratio Tokamaks ($\rho/R_0 \ll 1$) take place in the Low Aspect Ratio Toroidal Plasmas ($\rho/R_0 < 1$) with circular [16-18], elliptic [19], and D-shaped magnetic surfaces [20]. In both the Large and Low Aspect Ratio Tokamaks with circular magnetic surfaces, the equilibrium magnetic field has only one minimum (or three extremums, with respect to the poloidal angle $\theta$). Accordingly, in such plasma models there is only one group of trapped particles. The main feature of a toroidal plasma with elliptic magnetic surfaces is the fact that the equilibrium magnetic field can have two local minimums (or five extremums, with respect to $\theta$). As a result, together with untrapped and usual *t*-trapped particles, two additional groups of the so-called *d*-trapped (or double-trapped) particles can appear at such magnetic surfaces where the corresponding criterion [19] is satisfied: $\rho/R_0 < h_\theta^2(b^2/a^2 - 1)$, here *b/a* is the elongation, and $h_\theta$ is the poloidal component of the unit vector along the equilibrium magnetic field.

Many present-days tokamaks, mainly the spherical ones, have the D-shaped transverse cross-sections of the magnetic surfaces. In this paper, the parallel dielectric permittivity elements are derived for RF waves in an axisymmetric D-shaped toroidal plasma with a Solov'ev type equilibrium. A collisionless plasma model is considered. Drift-kinetic equation is solved separately for untrapped and three groups of trapped particles as a boundary-value problem, using an approach developed for low aspect ratio tokamaks with concentric circular [18], elliptic [19] and D-shaped [20] magnetic surfaces.

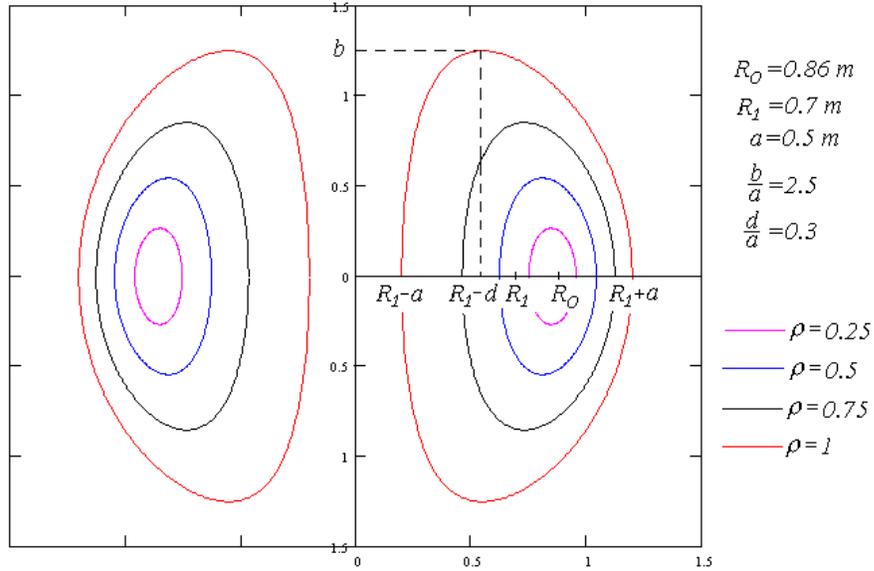

*Fig. 1. D-shaped transverse magnetic surface cross-sections in a tokamak with Solov'ev type equilibrium*

## 2. Plasma Model with Solov'ev Type Equilibrium

D-shaped transverse magnetic surface cross-sections corresponding to the Solov'ev type equilibrium [21, 22], see Fig.1, can be plotted by the following parametric equations for cylindrical $(R,\phi,Z)$ and quasi-toroidal coordinates $(\rho,\theta,\phi)$:

$$\boxed{\begin{aligned} R &= \sqrt{R_0^2 + 2aR_1 \rho \cos\theta} \\ \phi &= \phi \\ Z &= \frac{-\alpha\, aR_1 \rho\, \sin\theta}{\sqrt{R_0^2 - \delta + 2aR_1 \rho \cos\theta}} \end{aligned}} \qquad \boxed{\begin{aligned} \rho &= \frac{\sqrt{4Z^2(R^2 - \delta) + \alpha^2(R^2 - R_0^2)^2}}{2aR_1 \alpha} \\ \theta &= -\arctan\left(\frac{2Z\sqrt{R^2 - \delta}}{\alpha(R^2 - R_0^2)}\right) \\ \phi &= \phi \end{aligned}} \qquad (1)$$

where $R_0$ is the radius of the main magnetic axis;
$R_1$ is the major radius of the external magnetic surface;
$a$ is the (minor) plasma radius;
$b/a$ is the elongation;
$d/a$ is the triangularity;
$\rho$ is the non-dimensional radius of a local magnetic surface ($0 < \rho < 1$);
$\theta$ is the poloidal angle ($-\pi < \theta < \pi$);

and the additional definitions are [22]

$$\delta = \frac{(R_1 - a)^2 (R_1 + a)^2 - (R_1 - d)^4}{2(R_0^2 - (R_1 - d)^2)} ; \qquad \alpha^2 = \frac{2b^2}{R_0^2 - (R_1 - d)^2}. \tag{2}$$

The force balance and Maxwell's equations:

$$\mathbf{J} \times \mathbf{H} = c \nabla p, \qquad \nabla \times \mathbf{H} = (4\pi/c)\mathbf{J}, \qquad \nabla \cdot \mathbf{H} = 0, \tag{3}$$

can be reduced to the Grad-Shafranov equation [23, 24]

$$R \frac{\partial}{\partial R} \frac{1}{R} \frac{\partial \Psi}{\partial R} + \frac{\partial^2 \Psi}{\partial Z^2} = -\frac{4\pi}{c} R J_\phi = -4\pi R^2 \frac{\partial p}{\partial \Psi} - I_p \frac{\partial I_p}{\partial \Psi}, \tag{4}$$

where $J_\phi$ is the toroidal current density; $c$ is the speed of light; $p = p(\Psi)$ is the plasma pressure; $\Psi = R A_\phi$ is the poloidal magnetic flux; i.e., $\mathbf{H} = H_\phi \mathbf{e}_\phi + \nabla \times (A_\phi \mathbf{e}_\phi)$ and $I_p = I_p(\Psi) = R H_\phi$. The Solov'ev solution [21, 22] of the Grad-Shafranov equation has been found for the case

$$p = p_0 - \frac{\sqrt{p_0(1+\alpha^2)}}{a R_1 \alpha \sqrt{2\pi}} \Psi, \qquad I_p^2 = R_0^2 H_{0\phi}^2 + \frac{4\delta \sqrt{2\pi p_0}}{a R_1 \alpha \sqrt{1+\alpha^2}} \Psi, \tag{5}$$

and can be written as

$$\Psi = \frac{\sqrt{2\pi p_0}}{a R_1 \alpha \sqrt{1+\alpha^2}} \left[ Z^2 (R^2 - \delta) + \frac{\alpha^2}{4} (R^2 - R_0^2)^2 \right] = \frac{a R_1 \alpha \sqrt{2\pi p_0}}{\sqrt{1+\alpha^2}} \rho^2, \tag{6}$$

where $p_0$ and $H_{0\phi}$ are, respectively, the plasma pressure and toroidal magnetic field on the axis ($R = R_0$ or $\rho = 0$). Accordingly, the non-dimensional radius $\rho$, see Eqs. (1), is introduced as

$$\rho = \sqrt{\frac{\Psi \sqrt{1+\alpha^2}}{a R_1 \alpha \sqrt{2\pi p_0}}} = \frac{\sqrt{4Z^2(R^2 - \delta) + \alpha^2 (R^2 - R_0^2)^2}}{2 a R_1 \alpha}. \tag{7}$$

As a result, the components of an equilibrium magnetic field can be written as

$$H_R = \frac{1}{R} \frac{\partial \Psi}{\partial Z} = -\sqrt{\frac{2 p_0}{1+\alpha^2}} \rho \sin\theta \sqrt{\frac{R_0^2 - \delta + 2 a R_1 \rho \cos\theta}{R_0^2 + 2 a R_1 \rho \cos\theta}}$$

$$H_\phi = \frac{I_p}{R} = \frac{1}{R} \sqrt{R_0^2 H_{0\phi}^2 + \frac{2\delta\sqrt{2\pi p_0}}{a R_1 \alpha \sqrt{1+\alpha^2}} \Psi} = H_{0\phi} \sqrt{\frac{1 + \frac{\beta_o \delta \rho^2}{R_0^2(1+\alpha^2)}}{1 + 2 \frac{a R_1}{R_0^2} \rho \cos\theta}} \tag{8}$$

$$H_Z = -\frac{1}{R} \frac{\partial \Psi}{\partial R} = -\sqrt{\frac{2 p_0}{1+\alpha^2}} \alpha \rho \frac{a R_1 \rho (1+\cos^2\theta) + (R_0^2 - \delta)\cos\theta}{R_0^2 - \delta + 2 a R_1 \rho \cos\theta}$$

and module of an equilibrium magnetic field is

$$H_0 = \sqrt{H_R^2 + H_Z^2 + H_\phi^2} = H_{0\phi} G(\rho,\theta) = H_{0\phi} \sqrt{\frac{1 + \frac{\beta_o \rho^2}{1+\alpha^2} g(\rho,\theta)}{1 + 2\frac{aR_1}{R_0^2}\rho\cos\theta}} \qquad (9)$$

where

$$g(\rho,\theta) = \sin^2\theta\left(1 + 2\frac{aR_1}{R_0^2}\rho\cos\theta\right) + \frac{\delta}{R_0^2}\cos^2\theta + \alpha^2 \frac{1 + 2\frac{aR_1}{R_0^2}\rho\cos\theta}{\left(1 + \frac{2aR_1\rho}{R_0^2 - \delta}\cos\theta\right)^2}\left[\cos\theta + \frac{aR_1\rho}{R_0^2 - \delta}\left(1 + \cos^2\theta\right)\right]^2,$$

$\beta_o = \frac{8\pi p_0}{H_{0\phi}^2}$ is the $\beta$–value (central *beta*) on the main magnetic axis ($R=R_0$ or $\rho=0$).

Now let us verify how many minimums can have $H_0(\rho,\theta)$ in dependence of the poloidal angle $\theta$ at the given by $\rho$ magnetic surfaces. The corresponding plots are present in Fig. 2 for tokamaks with elliptic ($b/a=2.5$, $d/a=0$, $\rho=0.5$) and D-shaped ($b/a=2.5$, $d/a=0.3$, $\rho=0.5$) magnetic surfaces. As shown in Fig.2, plasma tends to have $H_0(\rho,\theta)$–configurations with one minimum. Configurations with two minimums of $H_0(\rho,\theta)$ (and, accordingly, additional groups of the trapped particles) can appear in elongated tokamaks with a high elongation $b/a > 2$ and high central $\beta_o$. Of course, the preferable regimes-configurations are those with the one minimum of $H_0(\rho,\theta)$, since the new-additional groups of the trapped particles can provoke the additional plasma-wave instabilities, modify the transport processes and so on.

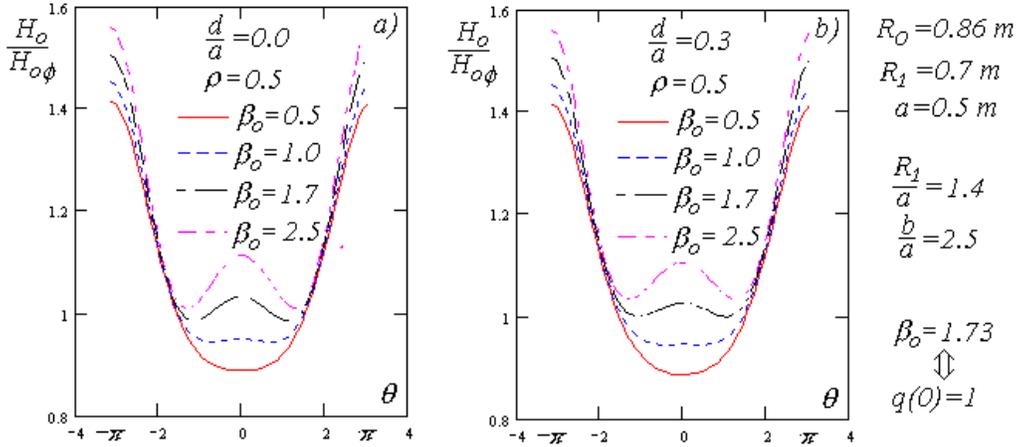

*Fig. 2. The dependence of the equilibrium magnetic field $H_0(\rho,\theta)$ on the poloidal angle $\theta$ in tokamaks with elliptic (a) and D-shaped (b) magnetic surfaces.*

Another important characteristic of tokamaks is a tokamak safety factor:

$$q(\rho) = \frac{q_o}{1 + 2\frac{aR_1}{R_0^2}\rho} \sqrt{\frac{1 + \frac{\beta_o \delta \rho^2}{R_0^2(1+\alpha^2)}}{1 + \frac{2aR_1\rho}{R_0^2 - \delta}}} \frac{2}{\pi}\Pi\left(\frac{\pi}{2}, \frac{4aR_1\rho}{R_0^2 + 2aR_1\rho}, \sqrt{\frac{4aR_1\rho}{R_0^2 - \delta + 2aR_1\rho}}\right) \qquad (10)$$

where $q_o = q(0) = \dfrac{aR_1\sqrt{1+\alpha^2}}{\sqrt{\beta_o}\,R_0\sqrt{R_0^2-\delta}}$ is the safety factor at the main magnetic axis ($R = R_0$ or $\rho=0$); and

$$\Pi(\eta,\lambda,\kappa) = \int_0^\eta \frac{d\varphi}{(1-\lambda\sin^2\varphi)\sqrt{1-\kappa^2\sin^2\varphi}} \qquad (11)$$

is the elliptic integral of the third kind. Note that $\beta_o$ and $q_o$ are self-consistent, i.e.,

$$\beta_o = \frac{a^2 R_1^2 (1+\alpha^2)}{q_o^2 R_0^2 (R_0^2-\delta)} . \qquad (12)$$

The radial structure of $q(\rho)$ in the low (*a*) and large (*b*) aspect ratio D-shaped tokamaks is present in Fig.3.

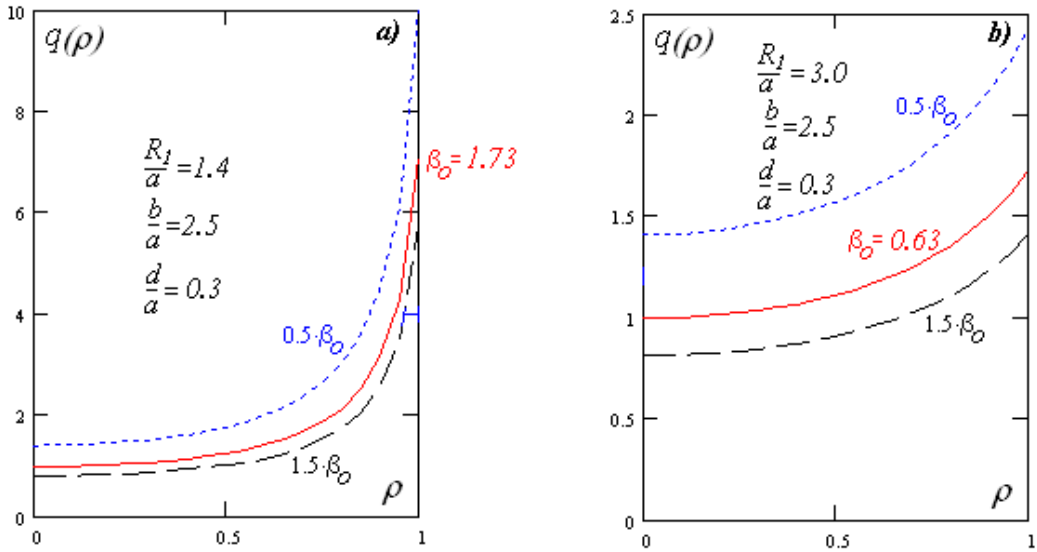

*Fig. 3. Radial structure of q(ρ) in the low (a) and large (b) aspect ratio tokamaks:*
*a) Low Aspect Ratio Tokamak  $R_1$=0.7 m, a=0.5 m, b=1.25 m, d=0.15 m;*
*b) Large Aspect Ratio Tokamak  $R_1$=3.0 m, a=1.0 m, b=2.5 m, d=0.3 m.*

### 3. Drift-Kinetic Equation

Using the conservation integrals $v_\parallel^2 + v_\perp^2 = const$ and $v_\perp^2/2H_0 = const$, the new variables $v$ and $\mu$ in velocity space can be introduced instead of $v_\parallel$ and $v_\perp$ as

$$v^2 = v_\parallel^2 + v_\perp^2, \qquad \mu = \frac{v_\perp^2}{v_\parallel^2 + v_\perp^2}\frac{1}{G(\rho,\theta)} . \qquad (13)$$

As a result, the drift-kinetic equation for the perturbed distribution functions,

$$f(\rho,\theta,v_\parallel,v_\perp) = \sum_{s=\pm 1} f_s(\rho,\theta,v,\mu)\exp(-i\omega t + in\phi) \qquad (14)$$

in the zeroth order over the magnetization parameter, can be reduced to

$$-i\omega f_s + s\frac{|v_\parallel|}{H_0}\left(H_R\frac{\partial\theta}{\partial R}+H_Z\frac{\partial\theta}{\partial Z}\right)\frac{\partial f_s}{\partial\theta}+ins\frac{|v_\parallel|H_\phi}{H_0 R}f_s = \frac{2e}{M}s\frac{|v_\parallel|}{v_T^2}E_\parallel F \quad (15)$$

where

$$H_0 = H_{0\phi}G(\rho,\theta), \qquad R = \sqrt{R_0^2 + 2aR_1\rho\cos\theta},$$

$$H_R\frac{\partial\theta}{\partial R}+H_Z\frac{\partial\theta}{\partial Z} = \frac{H_{0\phi}\sqrt{\beta_o}}{aR_1\sqrt{1+\alpha^2}}\sqrt{R_0^2-\delta+2aR_1\rho\cos\theta}, \quad (16)$$

$$F = \frac{N}{\pi^{1.5}v_T^3}\exp\left(-\frac{v^2}{v_T^2}\right), \qquad v_T^2 = \frac{2T}{M}.$$

By $s = \pm 1$ we distinguish the perturbed distribution functions $f_s$, with positive and negative values of the parallel velocity (relative to **H**) $v_\parallel = sv\sqrt{1-\mu\cdot G(\rho,\theta)}$.

## 4. Trapped and Untrapped Particles

The parallel current density components can be calculated as

$$j_\parallel(\rho,\theta) = \frac{\mathbf{j}\cdot\mathbf{H_0}}{H_0} = \pi eG(\rho,\theta)\sum_s^{\pm 1} s\int_0^\infty v^3\int_0^{1/G(\rho,\theta)} f_s(\rho,\theta,v,\mu)d\mu dv. \quad (17)$$

Analysing the conditions

$$v_\parallel(\mu,\theta) = sv\sqrt{1-\mu\cdot G(\rho,\theta)} = 0 \quad (18)$$

the phase volume of plasma particles should be split in the phase volumes of untrapped, *t*-trapped and *d*-trapped particles:

| | | |
|---|---|---|
| $0 \le \mu \le \mu_u$ | $-\pi \le \theta \le \pi$ | - for untrapped particles |
| $\mu_u \le \mu \le \mu_t$ | $-\theta_t \le \theta \le \theta_t$ | - for *t*-trapped particles |
| $\mu_t \le \mu \le \mu_d$ | $-\theta_t \le \theta \le -\theta_d$ | - for *d*-trapped particles |
| $\mu_t \le \mu \le \mu_d$ | $\theta_d \le \theta \le \theta_t$ | - for *d*-trapped particles |

where the reflection points $\pm\theta_t$ and $\pm\theta_d$ for *t*-trapped and *d*-trapped particles can be defined by solving the equation $G(\rho,\theta) = \mu$ and the extremums $\mu_u$, $\mu_t$ and $\mu_d$ are defined by

$$\mu_u = G(\rho,\pm\pi), \qquad \mu_t = G(\rho,0), \qquad \mu_d = G(\rho,\theta_{\max})$$

with the angles $\theta_{\max}$ satisfying the equation $dg(\rho,\theta)/d\theta = 0$. In particular,

$$\mu_u = \sqrt{\frac{1-2\frac{aR_1}{R_0^2}\rho}{1+\frac{\beta_o\rho^2}{1+\alpha^2}\left(\frac{\delta}{R_0^2}+\alpha^2\left(1-2\frac{aR_1}{R_0^2}\rho\right)\right)}}, \qquad \mu_t = \sqrt{\frac{1+2\frac{aR_1}{R_0^2}\rho}{1+\frac{\beta_o\rho^2}{1+\alpha^2}\left(\frac{\delta}{R_0^2}+\alpha^2\left(1+2\frac{aR_1}{R_0^2}\rho\right)\right)}}. \quad (19)$$

Using the boundary conditions
1) periodicity of $f_s$ over $\theta$ for *u*-particles;
2) continuity of $f_s$ at the stop-points $\pm\theta_t$ and $\pm\theta_d$ for *t*-trapped and *d*-trapped particles,

the perturbed distributions can be found in the form

$$f_s^u = \sum_p^{\pm\infty} f_{s,p}^u \exp\left[i2\pi(p+nq)\frac{\tau(\theta)}{T_u}-inq\bar\theta(\theta)\right], \quad (20)$$

$$f_s^{t,d} = \sum_p^{\mp\infty} f_{s,p}^{t,d} \exp\left[i2\pi p \frac{\tau(\theta)}{T_{t,d}} - inq\,\bar{\theta}(\theta)\right], \tag{21}$$

where $p$ is the number of the bounce resonances;

$$\tau(\theta) = \int_0^\theta \frac{G(\rho,\eta)\,d\eta}{\sqrt{1-\mu G(\rho,\eta)}\sqrt{1+\frac{2aR_1\rho}{R_0^2-\delta}\cos\theta}} \tag{22}$$

is the new time-like variable to describe the bounce-periodic motion of $u$-, $t$-, $d$-particles; the corresponding bounce periods are: $T_u = 2\tau(\pi)$, $T_t = 4\tau(\theta_t)$, $T_d = 2(\tau(\theta_t) - \tau(\theta_d))$; and

$$\bar{\theta}(\theta) = \pi \frac{\Pi\left(\frac{\theta}{2}, \frac{4aR_1\rho}{R_0^2 + 2aR_1\rho}, \sqrt{\frac{4aR_1\rho}{R_0^2 - \delta + 2aR_1\rho}}\right)}{\Pi\left(\frac{\pi}{2}, \frac{4aR_1\rho}{R_0^2 + 2aR_1\rho}, \sqrt{\frac{4aR_1\rho}{R_0^2 - \delta + 2aR_1\rho}}\right)} \tag{23}$$

is the new poloidal angle for coordinates where the magnetic field lines are 'straight'. The Fourier harmonics $f_{s,p}^u$, $f_{s,p}^t$ and $f_{s,p}^d$ can be derived after the corresponding bounce-averaging.

## 5. Parallel Permittivity Elements

To evaluate the parallel permittivity elements we use the Fourier expansions of the current density and electric field over $\bar{\theta}$:

$$\frac{j_\parallel(\theta)}{G(\rho,\theta)}\left(1 + \frac{2aR_1\rho}{R_0^2}\cos\theta\right)\sqrt{1 + \frac{2aR_1\rho}{R_0^2-\delta}\cos\theta} = \sum_m^{\pm\infty} j_\parallel^m \exp(im\bar{\theta}), \tag{24}$$

$$\frac{E_\parallel(\theta)G(\rho,\theta)}{\sqrt{1 + \frac{2aR_1\rho}{R_0^2-\delta}\cos\theta}} = \sum_{m'}^{\pm\infty} E_\parallel^{m'} \exp(im'\bar{\theta}). \tag{25}$$

As a result,

$$\frac{4\pi i}{\omega} j_\parallel^m = \sum_{m'}^{\pm\infty} \varepsilon^{m,m'} E_\parallel^{m'} = \sum_{m'}^{\pm\infty}(\varepsilon_{\parallel,u}^{m,m'} + \varepsilon_{\parallel,t}^{m,m'} + \varepsilon_{\parallel,d}^{m,m'})E_\parallel^{m'} \tag{26}$$

and the contributions of $u$-, $t$-trapped and $d$-trapped particles to $\varepsilon_\parallel^{m,m'}$ are

$$\varepsilon_{\parallel,u}^{m,m'} = \frac{\omega_p^2 R_0^2 q_o^2\left(1 + \frac{2aR_1\rho}{R_0^2}\right)\sqrt{1 + \frac{2aR_1\rho}{R_0^2-\delta}}}{4v_T^2\pi^2\Pi\left(\frac{\pi}{2}, \frac{4aR_1\rho}{R_0^2 + 2aR_1\rho}, \sqrt{\frac{4aR_1\rho}{R_0^2 - \delta + 2aR_1\rho}}\right)} \sum_{p=-\infty}^\infty \int_0^{\mu_u} \frac{T_u A_p^m A_p^{m'}}{(p+nq)^2}\left[1 + 2u_p^2 + 2i\sqrt{\pi}u_p^3 W(u_p)\right]d\mu$$

$$\varepsilon_{\parallel,t}^{m,m'} = \frac{\omega_p^2 R_0^2 q_o^2\left(1 + \frac{2aR_1\rho}{R_0^2}\right)\sqrt{1 + \frac{2aR_1\rho}{R_0^2-\delta}}}{4v_T^2\pi^2\Pi\left(\frac{\pi}{2}, \frac{4aR_1\rho}{R_0^2 + 2aR_1\rho}, \sqrt{\frac{4aR_1\rho}{R_0^2 - \delta + 2aR_1\rho}}\right)} \sum_{p=1}^\infty \int_{\mu_u}^{\mu_t} \frac{T_t}{p^2} B_p^m B_p^{m'}\left[1 + 2v_p^2 + 2i\sqrt{\pi}v_p^3 W(v_p)\right]d\mu$$

$$\varepsilon_{\parallel,d}^{m,m'} = \frac{\omega_p^2 R_0^2 q_o^2 \left(1 + \frac{2aR_1\rho}{R_0^2}\right)\sqrt{1 + \frac{2aR_1\rho}{R_0^2 - \delta}}}{2v_T^2 \pi^2 \Pi\left(\frac{\pi}{2}, \frac{4aR_1\rho}{R_0^2 + 2aR_1\rho}, \sqrt{\frac{4aR_1\rho}{R_0^2 - \delta + 2aR_1\rho}}\right)} \sum_{p=1}^{\infty} \int_{\mu_u}^{\mu_d} \frac{T_d}{p^2}(C_p^m C_p^{m'} + S_p^m S_p^{m'})\left[1 + 2z_p^2 + 2i\sqrt{\pi}z_p^3 W(z_p)\right]d\mu$$

Here we have used the following definitions:

$$u_p = \frac{\omega T_u R_0 q_0}{2\pi |p + nq_t| v_T}, \qquad v_p = \frac{\omega T_t R_0 q_0}{2\pi p v_T}, \qquad z_p = \frac{\omega T_d R_0 q_0}{2\pi p v_T}$$

$$A_p^m = \int_0^{\pi} \cos\left[(m + nq_t)\bar{\theta}(\eta) - (p + nq_t)2\pi \frac{\tau(\eta)}{T_u}\right] d\eta \qquad (27)$$

$$B_p^m = \int_0^{\theta_t} \cos\left[(m + nq_t)\bar{\theta}(\eta) - 2\pi p \frac{\tau(\eta)}{T_t}\right] d\eta + (-1)^{p+1} \int_0^{\theta_t} \cos\left[(m + nq_t)\bar{\theta}(\eta) + 2\pi p \frac{\tau(\eta)}{T_t}\right] d\eta$$

$$C_p^m = \int_{\theta_d}^{\theta_t} \cos[(m + nq_t)\bar{\theta}(\eta)]\sin\left[2\pi p \frac{\tau(\eta)}{T_d}\right] d\eta, \qquad \omega_p^2 = \frac{4\pi N e^2}{M}$$

$$S_p^m = \int_{\theta_d}^{\theta_t} \sin[(m + nq_t)\bar{\theta}(\eta)]\sin\left[2\pi p \frac{\tau(\eta)}{T_d}\right] d\eta, \qquad W(z) = \exp(-z^2)\left(1 + \frac{2i}{\sqrt{\pi}}\int_0^z \exp(t^2)dt\right).$$

Note that we have derived the contribution of any kind of *u*- and *t*-particles to the parallel permittivity elements. The corresponding expressions for plasma electrons and ions can be obtained replacing *T, N, M, e* by the electron $T_e$, $N_e$, $m_e$, $e_e$ and ion $T_i$, $N_i$, $M_i$, $e_i$ parameters, respectively. To obtain the total expressions of the dielectric permittivity elements, as usual, it is necessary to carry out the summation over the all species of plasma particles.

### 6. Wave Dissipation by Electron Landau Damping

One of the main mechanisms of the RF plasma heating is the electron Landau damping of waves due to the Cherenkov resonance interaction of $E_\parallel$ with the trapped and untrapped electrons. Here we should take into account that the Cherenkov resonance conditions are different for trapped and untrapped particles in tokamak plasmas and have nothing in common with the wave-particle resonance conditions in the cylindrical magnetised plasmas. Another important feature of the 2D tokamak plasmas is the contributions of the all $E_\parallel^{m'}$-harmonics to the given $j_\parallel^m$-harmonic, see Eq. (26). As a result, after averaging in time and poloidal angle, the wave power absorbed by the trapped and untrapped electrons, $P = \text{Re}(E_\parallel \cdot j_\parallel^*)$ can be estimated by the expression

$$P = \frac{\omega}{8\pi} \sum_m^{\pm\infty} \sum_{m'}^{\pm\infty} \left(\text{Im}\varepsilon_{\parallel,u}^{m,m'} + \text{Im}\varepsilon_{\parallel,t}^{m,m'} + \text{Im}\varepsilon_{\parallel,d}^{m,m'}\right)\left(\text{Re}E_\parallel^m \text{Re}E_\parallel^{m'} + \text{Im}E_\parallel^m \text{Im}E_\parallel^{m'}\right), \qquad (28)$$

where $\text{Im}\varepsilon_{\parallel,u}^{m,m'}$, $\text{Im}\varepsilon_{\parallel,t}^{m,m'}$ and $\text{Im}\varepsilon_{\parallel,d}^{m,m'}$ are the contributions of untrapped, *t*-trapped and *d*-trapped electrons to the imaginary part of the parallel permittivity elements.

### Conclusion

Our dielectric characteristics can be applied for both the large and low aspect ratio tokamaks with circular, elliptic and D-shaped magnetic surfaces to study the wave processes with a regular frequency such as the wave propagation and wave dissipation during the plasma heating and current drive generation; when the wave frequency has been done, e.g., by the antenna-generator system, in the frequency range of the Alfvén and fast magnetosonic waves.

Analysing the collisionless wave dissipation by plasma particles we should remember the other kinetic mechanisms of the wave-particle interactions, such as TTMP (Transit Time Magnetic Pumping) and cyclotron resonance damping. Those can be described by the transverse and cross-off dielectric permittivity elements. A comprehensive theoretical analysis of the kinetic wave dissipation should be developed using all nine dielectric tensor components, accounting for the finite particle-orbit widths, the finite beta, and the finite Larmor radius effects. These corrections can be derived by solving the Vlasov equation for trapped and untrapped particles in the next approximation(s) over the magnetisation parameter. However, this is a topic for additional investigation.

*Acknowledgements:* This research was supported by CNPq of Brazil (Conselho Nacional de Desenvolvimento Científico e Tecnológico, projeto PCI-LNCC/MCT 382042/04-2).

**References**
[1] Y-K.M. Peng, D.J. Strickler, *Nuclear Fusion*, 26, 769, 1986.
[2] D.C. Robinson, *Fusion Energy and Plasma Physics*, World Scientific Press, 1987.
[3] A. Sykes *et al.*, *Plasma Phys. Control. Fusion*, 39 (B), 247, 1997.
[4] S.M. Kaye *et al.*, *Phys. Plasmas*, 8, 1977, 2001.
[5] G.F. Counsell *et al.*, *Plasma Phys. Control. Fusion*, 44, 23, 2002.
[6] V.K. Gusev, F. Alladio and A. W. Morris, *Plasma Phys. Control. Fusion,* 45, A59-A82, 2003
[7] T. Stix, 1975, *Nucl. Fusion,* 15, 737, 1975.
[8] V.N. Belikov, Ya.I. Kolesnichenko, A.B. Mikhailovskii, V.A. Yavorskii, *Sov. J. Plasma Phys.,* 3, 146, 1977
[9] T.D. Kaladze, A.I. Pyatak, K.N. Stepanov, *Sov. J. Plasma Phys.,* 8, 467, 1982.
[10] M. Brambilla, T. Krucken, *Nucl. Fusion,* 28, 1813, 1988.
[11] A.G. Elfimov, S. Puri, *Nucl. Fusion,* 30, 1215, 1990.
[12] F.M. Nekrasov, *Sov. J. Plasma Phys.*, 18, 520, 1992.
[13] F. Porcelli, R. Stancievicz, W. Kerner, H.L. Berk, *Phys. Plasmas*, 1, 470, 1994.
[14] S.V. Kasilov, A.I. Pyatak, K.N. Stepanov, *Plasma Phys. Reports*, 24, 465, 1998.
[15] B.N. Kuvshinov, A.B. Mikhailovskii, *Plasma Phys. Reports*, 24, 623, 1998.
[16] N.I. Grishanov, F.M. Nekrasov, 1990, *Sov. J. Plasma Phys.*, 16, 129, 1990.
[17] F.M. Nekrasov, A.G. Elfimov, C.A. de Azevedo, A.S. de Assis, *Plasma Phys. Control. Fusion*, 43, 727, 2001.
[18] N.I. Grishanov, C.A. de Azevedo, J.P. Neto, *Plasma Phys. Contr. Fusion*, 43,1003, 2001.
[19] N.I. Grishanov, G.O. Ludwig, C.A. de Azevedo, J.P. Neto*, Phys. Plasmas,* 9, 4089, 2002.
[20] N.I. Grishanov, A.F.D. Loula, C.A. de Azevedo, J.P. Neto*, Plasma Phys. Control. Fusion*, 45, 1791, 2003.
[21] L.S. Solov'ev*, Sov. Phys. JETP,* 26, 400, 1966.
[22] K. Yamazaki, *Preprint IPPJ-413*, Nagoya University, 26 p., 1979
[23] V.D. Shafranov, *Sov. Phys. JETP,* 6, 545, 1957.
[24] H. Grad, H. Rubin, *2nd UN Conference on Peaceful Uses of Atomic Energy,* 1958.